\documentclass[twocolumn, longbibliography]{revtex4-2}

\usepackage{amsmath}
\usepackage[dvipsnames]{xcolor}
\usepackage{hyperref}
\hypersetup{
	colorlinks=true,
	linkcolor=blue,   
	urlcolor=blue,
}
\usepackage{graphicx}

\renewcommand{\l}{\ell}
\newcommand{\C}{\mathcal{C}}

\newcommand{\TC}{\text{TC}}

\newcommand{\nn}{\nonumber\\}

\newcommand{\tdtf}{\text{3d3f}}

\newcommand{\para}{\text{para}}
\renewcommand{\vert}{\text{vert}}
\newcommand{\horiz}{\text{horiz}}

\newcommand{\pardag}{\partial^\dagger\!}
\newcommand{\ket}[1]{\left|#1\right\rangle}

\begin{document}
	
\title{Symmetry protected self correcting quantum memory in three space dimensions}
\author{Charles Stahl}
\author{Rahul Nandkishore}
\affiliation{Department of Physics and Center for Theory of Quantum Matter, University of Colorado Boulder, Boulder CO 80309, USA}

\begin{abstract}
Whether self correcting quantum memories can exist at non-zero temperature in a physically reasonable setting remains a great open problem. It has recently been argued \cite{RobertsBartlett} that symmetry protected topological (SPT) systems in three space dimensions subject to a strong constraint---that the quantum dynamics respect a {\it 1-form symmetry}---realize such a quantum memory. We illustrate how this works in Walker-Wang codes, which provide a specific realization of these desiderata. In this setting we show that it is sufficient for the 1-form symmetry to be enforced on a sub-volume of the system which is measure zero in the thermodynamic limit. This strongly suggests that the `SPT' character of the state is not essential. We confirm this by constructing an explicit example with a {\it trivial} (paramagnetic) bulk that realizes a self correcting quantum memory. We therefore show that the enforcement of a 1-form symmetry on a measure zero sub-volume of a three dimensional system can be sufficient to stabilize a self correcting quantum memory at non-zero temperature. 
\end{abstract}
	
\date{\today}
	
\maketitle
	
\section{Introduction}
A self correcting quantum memory can robustly store quantum information without need for active error correction, because its native dynamics suppresses errors for a time that diverges in the thermodynamic limit. The toric code in four space dimensions~\cite{Kitaev2003} provides a paradigmatic example of a self correcting quantum memory, in which the self correction property survives to non-zero temperature. However, whether these desirable properties can be realized in a physically reasonable system remains a great open problem. As far as we are aware, no such examples are currently known. `Fracton' models like the Haah cubic code \cite{HaahCode} come close, but alas, at non-zero temperature the memory time saturates to some temperature dependent finite value, even in the thermodynamic limit \cite{Siva2017, PremHaahNandkishore}. 

Roberts and Bartlett (R\&B) have recently shown \cite{RobertsBartlett} that a symmetry enriched topological phase on the two dimensional boundary of a three dimensional symmetry protected topological (SPT) bulk can realize a self correcting quantum memory at non-zero temperature, if we enforce a strong constraint---namely that the dynamics respects a {\it 1-form symmetry}. A 1-form symmetry \cite{Gaiotto2015, LakeHigher, Tong2018, WenHigher, Qi2021} is a symmetry that acts on manifolds of co-dimension one, and thus represents a very strong constraint. This remarkable breakthrough serves as the inspiration for the present work. 

In this article we show how a self correcting quantum memory may be realized in Walker-Wang models, thereby extending the R\&B construction to a new family of models. Additionally, we point out in this context that it is sufficient for the 1-form symmetry to be enforced in a volume which is measure zero in the thermodynamic limit. This strongly suggests that it is inessential for the bulk to be in an SPT phase. We confirm this by constructing an example whereby enforcement of a 1-form symmetry gives rise to a self correcting quantum memory on the boundary of a {\it trivial} three dimensional bulk. Thus we show that enforcement of a 1-form symmetry in a measure zero sub-volume of a three dimensional system can be sufficient to give rise to self correction. 

To set the stage for the discussion it is useful to first review the well known physics of toric codes \cite{Kitaev2003}. The ground states of the 2d toric code are loop gasses, in that they can be written as a product of closed loop operators acting on a reference state. The ground states of the 4d toric code are membrane condensates in the same sense. We will refer to these loops and membranes as nonlocal stabilizers, because they are elements of the stabilizer group that have a large support. All elements of the stabilizer group, both local and nonlocal, are closed. One characteristic of topological order is a ground-state degeneracy on manifolds with non-trivial topology. Nontrivial operators on the ground space are non-contractible versions of the nonlocal stabilizers. In the 2d toric code they are non-contractible loops, while in the 4d toric code they are non-contractible membranes. Excitations above the ground state appear at the boundaries of open versions of nonlocal stabilizers. For the 2d toric code these are point-like excitations on the ends of strings, while in the 4d toric code they are flux tube-like excitations on the boundary of open membranes. Finally, the 3d toric code has one sector with stabilizers that look like those of the 2d toric code and one with stabilizers that look like those of the 4d toric code. As such, its ground states can be written as loop gasses or membrane gasses.
	
All three toric codes have topological order at zero temperature, but have different nonzero temperature behavior. In four dimensions the toric code remains (quantum) topologically ordered up to some transition temperature $T_*>0$, while the 2d toric code is trivially ordered for any nonzero temperature. The 3d toric code remains topologically ordered for small nonzero temperatures, but the order is classical~\cite{CastelnovoFiniteTemp}. From the information theory perspective this means the code can protect a classical probabilistic bit 
but not a qubit. 
	
In both the 2d and 3d toric codes the nonzero temperature behavior can be traced to the finite energy barrier $\Delta<\infty$. The bath can lend a constant amount of energy to create two point defects and then transport them at no energy cost across the system. When they annihilate they leave behind a non-contractible nonlocal stabilizer, which we said acted nontrivially on the ground space. For the 4d toric code, the bath must create a membrane that stretches across the system. Since the energy cost of open membranes is linear in perimeter, the energy barrier to membrane operators is linear in system size. In the thermodynamic limit the energy barrier $\Delta$ is unbounded.

With this motivation, considerable work has been done to try to find 3-dimensional systems with unbounded energy barriers, and a number have been found, such as Haah's cubic code~\cite{HaahCode} and Michnicki's welded code~\cite{MichnickiPowerLaw}. They are collectively referred to as marginally self-correcting~\cite{Siva2017}. These codes have an energy barrier that grows less than linearly, either logarithmically (Haah's) or polynomially (Michnicki's). However, it has been shown that the bath still disorders these models at any $T>0$, so that the memory time is bounded independent of system size \cite{Siva2017, PremHaahNandkishore}. As in the 2d and 3d toric codes, the marginally self-correcting models have point-like excitations. At nonzero temperature these excitations exist at some nonzero density, leading to an energy barrier that is bounded by a function of the temperature.

The R\&B proposal directly removes the point excitations from the picture. This is achieved by enforcing what is called a 1-form symmetry ~\cite{Gaiotto2015, LakeHigher}.  Enforcing the symmetry is equivalent to giving the relevant Hamiltonian terms infinite coupling constants.
For example, consider the 2d toric code. If the dynamics are restricted to states where the plaquette and vertex terms have eigenvalue $+1$, then no point excitations can exist. This is an example of an enforced 1-form symmetry, defined in Sec.~\ref{sub:1form}. However, this is not an example of a self correcting quantum memory, because the logical operators cannot be applied transversally, i.e. as a series of local operations which respect the symmetry. Thus, enforcing a 1-form symmetry on the 2d toric code eliminates the pointlike excitations, but at the cost of our ability to apply logical operators. 

The R\&B proposal \cite{RobertsBartlett} instead creates a code that, when the symmetry is enforced, behaves like the 4d toric code in that logical operators can be applied transversally but with a large enough energy barrier that the bath applies them with probability 0 in the thermodynamic limit, at sufficiently low but non-zero temperature. This is achieved using a 2d topological order on the boundary of a 3d SPT. 
	
In this paper we show how to achieve the same results using the 3d 3-fermion model~\cite{BurnellSoluble}, a specific example of a confined Walker-Wang model. We expect that this prescription should work for any confined Walker-Wang model~\cite{WalkerWang, vonKeyserlingkSurfaceAnyons}. We show that the relevant symmetry need only be enforced ``close" to the boundary, in a sense that we will explain, suggesting that the SPT nature of the bulk may be inessential to the phenomenon. 
We then show that a model with a trivial paramagnetic bulk can display the same phenomena. The 1-form symmetry directly protects the quantum memory by introducing an appropriate coupling between pointlike excitations on the boundary and confined fluxes in the bulk. 
We conclude with a discussion of 1-form symmetry protection in the topologically ordered 3d toric code and some discussion of possible future work.

\section{Self-correction in the three-fermion model}
	
In this section we will first define the 3d 3-fermion model in the absence of the protecting symmetry and show it is not self-correcting. We then define the 1-form symmetry and show what nonlocal stabilizers and excitations can exist in its presence. Finally, we show the 3d 3-fermion model is self-correcting in the presence of the 1-form symmetry.

Confined Walker-Wang models---such as the 3d 3-fermion model---are a natural setting for this procedure. Like the model in Ref.~\cite{RobertsBartlett}, they describe 2d topological order on the boundary of a 3d trivial bulk. As the name suggests, they can be interpreted as models where anyons are deconfined on the boundary and confined by a linear potential in the bulk. We will see that the 1-form symmetry forces any anyons traveling across the boundary to be connected to anyons traveling through the bulk. Linear confinement in the bulk is then what gives this model an unbounded energy barrier.
	
\subsection{The model}

The three-fermion model can be viewed as two copies of the 3d toric code, ``twisted" together so that flux from one code confines the point-like excitations of the other. To be concrete, consider a cubic lattice with two qubits on each edge. We will refer to them as $\sigma$ and $\tau$ qubits, and they will be acted on by Pauli matrices written as $\sigma^\alpha$ and $\tau^\alpha$ respectively, with $\alpha = x,z$. Two independent toric codes would have the Hamiltonian 
\begin{align}
H_{\TC} &= -\sum_vA_v^{\sigma}-\sum_vA_v^{\tau}-\sum_fB_f^{\sigma0}-\sum_fB_f^{\tau0},\nn
A_v^{\sigma} &= \prod_{e\in \pardag v}\sigma_e^x, \qquad B_f^{\sigma0} = \prod_{e\in\partial p}\sigma_e^z,\nn
A_v^{\tau} &= \prod_{e\in \pardag v}\tau_e^x, \qquad B_f^{\tau0} = \prod_{e\in\partial p}\tau_e^z, \label{eqn:toric}
\end{align}
so that the two codes do not talk to each other at all. We will refer to the two types of terms as vertex terms and face terms. Here $\partial$ is the boundary operator and $\pardag$ is the dual boundary operator. These operators are related in that $a\in\partial b$ is equivalent to $b\in\pardag a$.
	
In each code there are string-like operators with point-like excitations and membrane operators with loop-like excitations. We will call flipped $A_v^\sigma$ terms $e$-particles and flipped $A_v^\tau$ terms $m$-particles. Since flipped $B_f^\sigma$ and $B_f^\tau$ terms naturally come in dual lines, we will refer to them as $\sigma$-flux and $\tau$-flux, respectively. $e$-particles see $\sigma$-flux (with non-trivial Berry phase) and $m$-particles see $\tau$-flux. Finally, $e$-particles exist on the ends of $e$-strings, $\sigma$-flux lives on the boundaries of $\sigma$-membranes, etc.

\begin{figure}
\centering
\includegraphics[width=\linewidth]{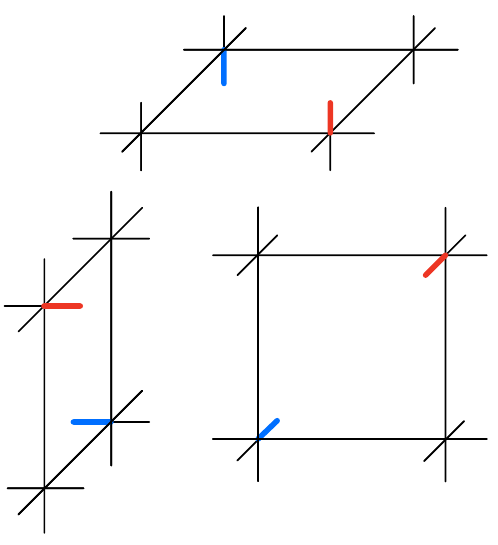}
\caption{Once we have fixed a projection, we can choose the $O$ and $U$ legs to be the ones the lie over and under the plaquette. In this illustration the $O$ legs are red and the $U$ legs are blue.}
\label{fig:legs}
\end{figure}

We now twist the codes together by decorating the face operators to create the 3d 3-fermion Hamiltonian,
\begin{align}
H_{\tdtf} &= -\sum_vA_v^{\sigma}-\sum_vA_v^{\tau}-\sum_fB_f^{\sigma}-\sum_fB_f^{\tau},\nn
A_v^{\sigma} &= \prod_{e\in \pardag v}\sigma_e^x, \qquad B_f^{\sigma} = \sigma_O^x\sigma_U^x\tau_U^x\prod_{e\in\partial p}\sigma_e^z,\nn
A_v^{\tau} &= \prod_{e\in \pardag v}\tau_e^x, \qquad B_f^{\tau} = \sigma^x_O\tau^x_O\tau^x_U\prod_{e\in\partial p}\tau_e^z, \label{eqn:3d3f}
\end{align}
where the edges O and U lie ``over" and ``under" the given face, given a specific choice of 2d projection. This is shown in Fig.~\ref{fig:legs}, where the $O$ edges are red and the $U$ edges are blue. We will see that the result of this decoration is that, for example, a string of $\sigma^z$ operators that would usually create two deconfined $e$ particles now also creates a line of $\tau$-flux and two lines of $\sigma$-flux. This means point excitations are confined in the bulk.
	
Membrane operators are the same as they were in the toric code, being dual membranes of $\sigma^x$ or $\tau^x$ operators. However, a ``bare" string operator consisting of $\sigma^z$ or $\tau^z$ now creates flux excitations along its entire length in addition to creating point excitations on its ends. In particular, a bare string of $\sigma^z$ operators creates two lines of $\sigma$-flux and one line of $\tau$-flux. A bare string of $\tau^z$ operators creates two lines of $\tau$-flux and one line of $\sigma$-flux.

\begin{figure}
\centering
\includegraphics{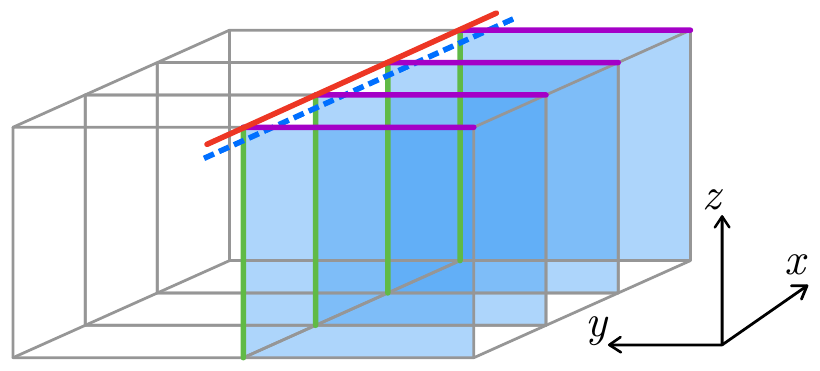}
\caption{In order to define the $S_\C^e$ on the red line $\C$, first draw the line $\C'$, which is the dashed blue line. Then the ``over" decoration is the purple legs and the ``under'' decoration is the green legs. In the end, the shaded blue faces are $\tau$-flux. Compare to Fig.~4 in Ref.~\cite{BurnellSoluble}.}
\label{fig:bulkstring}
\end{figure}

Since this is a model of $\mathbf{Z}_2$ topological order, the two lines of $\sigma$-flux that a string of $\sigma^z$ operators makes can be locally removed. Explicitly, we can construct the decorated string operator
\begin{align}
S^e_\mathcal{C} = \prod_{j\in\text{under}} \tau^x_j \sigma^x_j \prod_{k\in\text{over}} \sigma^x_k \prod_{i\in\mathcal{C}}\sigma^z_i,
\end{align}
where $\mathcal{C}$ is a curve, possibly open. To understand the decorations first draw a line $\mathcal{C}'$ that is equal to $\mathcal{C}$ offset infinitesimally in the $+\hat{x}-\hat{y}-\hat{z}$ direction (note this is a different direction than in~\cite{BurnellSoluble} because our axes are aligned differently and we will access a different boundary). The decoration ``over" consists of all edges adjacent to $\C$ that lie over $\C'$ (in our 2d projection), while the decoration ``under" consists of edges adjacent to $\C$ that lie under $\C'$. This configuration is shown in Fig.~\ref{fig:bulkstring}. 

The entire configuration leaves behind a string of $\tau$-flux, which cannot be locally removed. We therefore find that $e$-particles, which are created at the endpoints of $S^e_\C$, are linearly confined in the bulk.

There is also an $S^m_\mathcal{C}$ operator,
\begin{align}
S^m_\mathcal{C} = \prod_{j\in\text{under}} \tau^x_j  \prod_{k\in\text{over}} \tau^x_k \sigma^x_j \prod_{i\in\mathcal{C}}\tau^z_i,
\end{align}
which creates $m$-particles at its endpoints. It also leaves behind a single line of $\sigma$-flux, so the $m$-particles are also confined. Finally, there is a composite operator $S^\epsilon_\C=S^e_\C S^m_\C$ that creates composite $\epsilon$ particles confined by composite flux.
	
The flux that confines the point particles is the same as the flux on the boundary of membranes, in that both are dual lines of flipped face operators. We can then view the decorations on the string operator as a long narrow membrane whose boundary excitations cancel the superfluous flux lines. However, for both $S^e_\mathcal{C}$ and $S^m_\C$ there is one line of flux that cannot be canceled.

Confinement means the 3d 3-fermion model contains no topological order in the bulk, because there is no way to transport point particles across the system and return to the ground space. The result is that the 3d 3-fermion model is trivial when defined on manifolds without boundary.
	
\begin{figure}
\centering
\includegraphics[width=.6\linewidth]{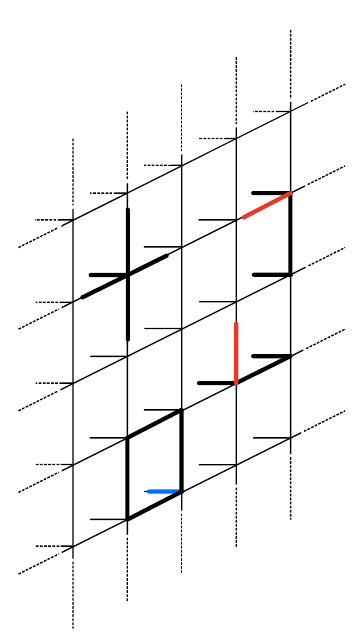}
\caption{The stabilizers on the boundary are truncated versions of the ones in the bulk. Red represents $O$ edges and blue represents $U$ edges. The two face operators that reach into the bulk each have a $U$ edge that is not shown (they are not truncated), while the boundary face operator does not have any $O$ edge.}
\label{fig:bdyops}
\end{figure}

On a manifold with a boundary, it is easy to terminate the code in a way that creates topological order. To do this, truncate the lattice using ``smooth" boundary conditions, so that no legs are sticking out. Then truncate any stabilizers to include all their operators that act on qubits that haven't been removed. Such stabilizers are shown in Fig.~\ref{fig:bdyops} The result is a 2d $\mathbf{Z}_2$ topological order where all anyons are fermions~\cite{BurnellSoluble}.

We emphasize that this is a choice of boundary conditions. It is possible to add a 2d 3-fermion model to the boundary and condense pairs, removing the topological order. However, since the boundary order is topological, it cannot be removed by arbitrarily small perturbations. Furthermore, it is possible to protect the boundary topological order by enforcing a 0-form time reversal symmetry~\cite{BurnellSoluble}. In that sense the bulk is SPT-ordered. We will instead enforce a 1-form symmetry, as described in the next subsection.
	
\begin{figure}
\centering
\includegraphics{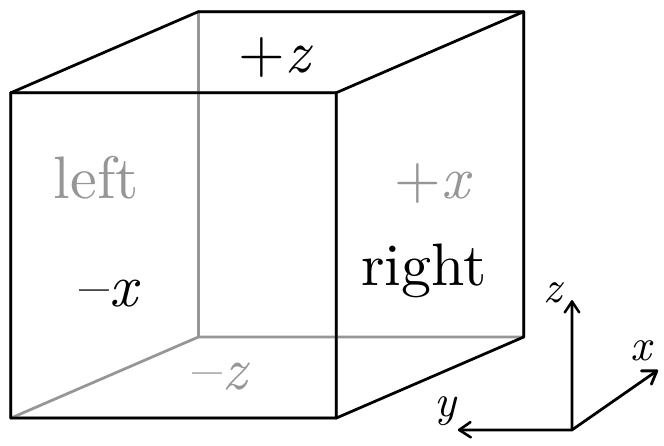}
\caption{Orientation for the $T^2\times I$ 3d 3-fermion model. The $\pm z$ sides are identified and the $\pm x$ sides are identified. The ``left" and ``right" boundaries at $y=1$ and $y=0$ respectively both have the topology of a torus. Both boundaries support two logical qubits.}
\label{fig:topo}
\end{figure}

We will consider the 3d 3-fermion model defined on a lattice with topology $T^2\times I$, where $T^2$ is the torus and $I$ is the unit interval $[0,1]$. This can be constructed from a cubic lattice by identifying the boundaries in the $x$- and $z$-directions, so that the only true boundaries are at $y=0,1$. We will refer to these as the the right and left boundaries, respectively. Each boundary supports two qubits. This configuration can be found in Fig.~\ref{fig:topo}.

We could call the topology $T^2\times I$ the hollow donut, because it can be embedded in flat 3d space by taking the core out of a solid donut. Then the two boundaries are the inner and outer boundary. Both boundaries have the topology of a plain old 2-torus.

If the topological order exists on the boundary, there must be logical operators supported only on boundary qubits. For the right boundary these are the deconfined string operators
\begin{align}
S^e_\C &= \prod_{j\in \text{under}}\tau_j^x \sigma_j^x \prod_{i\in\mathcal{C}} \sigma_i^z,\\
S^m_\C &= \prod_{j\in \text{under}}\tau_j^x\phantom{ \sigma_j^x} \prod_{i\in\mathcal{C}} \tau_i^z,
\end{align}
where $\mathcal{C}$ is now a line on the boundary. These are just truncated versions of the bulk operators. Only the ``under" legs get decorated because the ``over" legs have been removed from the lattice.

\begin{figure}
\centering
\includegraphics[width=.5\linewidth]{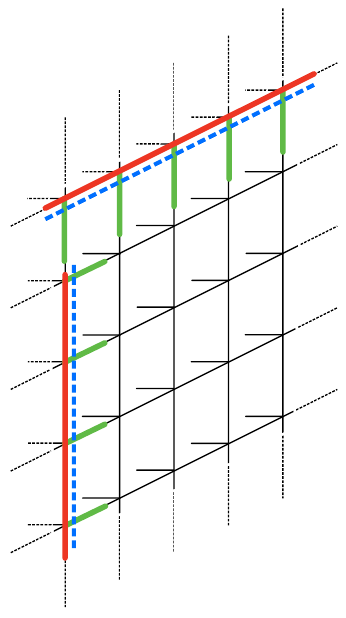}
\caption{The deconfined boundary string operators are truncated versions of the bulk string operators. The dashed blue line is once again $\C'$, but now there are no ``over" decorations. The green lines are the ``under" decorations.}
\label{fig:deconf}
\end{figure}

These operators create excitations at the endpoints of $\C$ but do not create flux along their length.  In fact, if we compare to Fig.~\ref{fig:bulkstring}, we see that the faces where the confining flux would exist have been removed from the lattice.
We can think of the flux as having been removed at the boundary by the decorations.  Fig.~\ref{fig:deconf} shows these decorations. 

Since there is no flux left, both types of string operators create deconfined anyons. Thus we have topological order.
If $\C$ is a noncontractible closed loop on the boundary, then the corresponding string operators are nontrivial operators on the ground space.

We will refer to a deconfined boundary string operator that wraps the vertical direction as $S_\vert$ and one that wraps the horizontal direction as $S_\horiz$. It is easy to check that 
\begin{align}
\{S^e_\vert,S^m_\horiz\}=\{S^e_\horiz,S^m_\vert\}=0
\end{align}
while all other pairs commute. Thus we could encode the logical operators as $Z_1=S^e_\vert$, $X_1 = S^m_\horiz$, $Z_2 = S^m_\vert$, and $X_2 = S^e_\horiz$, or any unitary transformation of that encoding. Similar string operators exist on the left boundary.

We previously mentioned the membrane operators in the model are the same as those in the 3d toric code. The membrane operators also appear as logical operators for the boundary topological order, with the caveat that they act nontrivially on both boundaries. For example there is 
\begin{align}
R^\sigma_\horiz = \prod_{i\in \mathcal{M}^*} \sigma^x_i,
\end{align}
which consists of $\sigma^x$ operators on every edge in a horizontal dual membrane $\mathcal{M}^*$. On the right boundary this acts as $X_1$, while it will also have a logical action on the left boundary.

The bath is able to transport deconfined point particles across the a system at any temperature above zero. This is the case in both the 2d and 3d toric code. In our case, all logical operators can be applied by transporting a deconfined point excitation across a boundary. 3D3F cannot store any information, even classical, at nonzero temperature. The same is true of confined Walker-Wang models in general. The topological order behaves the same as a 2d topological phase placed on the boundary of a trivial 3d bulk phase. We can however couple the boundary and the bulk using a higher-form symmetry as described below.

\subsection{Enforcing a 1-form symmetry} \label{sub:1form}

Here we define $p$-form symmetries, which for $p>0$ are called higher-form symmetries. A $p$-form symmetry consists of symmetry operators each associated with a closed $(d-p)$-dimensional submanifold of our space. The simplest examples, 0-form symmetries, are just ordinary global symmetries. They act on closed $d$-dimensional submanifolds, so they have to act on the whole space.
	
It may be unintuitive to think about symmetry operators that act on lower dimensional submanifolds. But toric codes actually provide convenient settings to think about them. In the 3d toric code, arbitrary products of vertex operators form (dual) membrane operators. These operators commute with the Hamiltonian, so they form a symmetry. The are defined on $(2=d-1)$-dimensional submanifolds, so they form a 1-form symmetry. 

We can write this symmetry group as $G=\left\langle A_v\right \rangle$, which means that $G$ is the group generated by all the $A_v$ operators.
The face terms form a 2-form symmetry $G'=\langle B_f\rangle$, but we are not concerned with that here.

Since the vertex terms were not affected when we twisted our toric codes together, the 3d 3-fermion model inherits the same 1-form symmetry. In particular, the symmetry group is 
\begin{align}
G = \langle A_v^\sigma\rangle \times \langle A_v^\tau \rangle,
\end{align}
the group generated by both types of vertex terms.

Recall that we wanted to get rid of point-like excitations on the boundary. We can do this by initializing the system in a state $\ket{\psi}$ that satisfies $g\ket{\psi}=\ket{\psi}$ for every $g$ in $G$. This includes the ground state and any state reached from the ground state by acting with open membrane operators. We then require that the dynamics obey the symmetry, so that no point particles are created. We will refer to this process as enforcing the symmetry $G$.

When we couple the system to a bath we can enforce $G$ by ensuring that all of the bath couplings commute with every element in $G$. This procedure is equivalent to giving $A_v^\sigma$ and $A_v^\tau$ infinite coupling constants. Enforcing the symmetry also prevents any open string operators. The symmetry still allows closed strings and open or closed membranes. 

Enforcing the symmetry $G$ ensures that every state in a local decomposition performed by the bath will respect the symmetry $G$. We will refer to this type of decomposition as a symmetric local decomposition~\cite{RobertsBartlett}.

Because the symmetry allows open membranes, any logical membrane operator can be decomposed into a series of local operations that do not break the symmetry. Logical string operators, on the other hand, must include open strings  in their local decompositions. This means that while logical string operators can be applied in the presence of the symmetry (because they are closed), they cannot be applied transversally without breaking the symmetry.

The ``problem" operators in the 3d 3-fermion model are the deconfined boundary string operators. 
Since the deconfined strings only exist on the boundary, it is tempting to only enforce the symmetry on the boundary. However, we can then create a string operator that lies mostly on the boundary but whose endpoints are in the bulk. Then the symmetry is only violated in the bulk, but the energy barrier is small.

If we enforce the symmetry in the bulk, then configurations that look like boundary anyons must be accompanied by bulk flux. Consider a closed string that intersects the boundary but is not entirely included in the boundary. Then on the boundary this looks like an open string that would create point excitations at its endpoints. However, at these   ``endpoints" the string instead goes into the bulk, where it is now confined and creates flux.

In this sense the 1-form symmetry couples bulk excitations to boundary excitations. This perspective will become most clear when we couple a 2d toric code to a paramagnet bulk in Sec.~\ref{sec:trivial}.

If the 1-form symmetry is enforced to a distance $W$ from the boundary, a nontrivial logical operator can be symmetrically decomposed into a series of strings whose endpoints are at least a distance $W$ from the boundary. 
In the following subsection we will define the symmetric energy barrier as the amount of energy the bath must provide in order to perform a logical operation. For the partially symmetry-protected 3d 3-fermion model it is $\Delta\sim W$.

\subsection{Diverging symmetric energy barrier} \label{sub:ener}
	
Since we assume the bath couples to the system locally, it can only apply a logical operator by decomposing it into a series of operators that differ by local operations. These operators generically create excitations in the system. Informally, the energy barrier is the energy of these excitations. We define the energy barrier more formally following Ref.~\cite{RobertsBartlett}.
	
First assume the bath couples to the system through local Pauli operators. Let $\bar{\l}$ be a (nontrivial) logical operator. Define the local decomposition of $\bar{\l}$ as a series of operators $\mathcal{D}(\bar{\l}) = \{\l^{(k)}| k = 1,\dots,N\}$, where $\l^{(1)}=I$ and $\l^{(N)}=\bar{\l}$. Furthermore, $\l^{(k)}$ and $\l^{(k+1)}$ differ only by a local (constant-range) set of Pauli operators. Since every Pauli operator either commutes or anti-commutes with each stabilizer, each of the $\l^{(k)}$ anticommutes with a finite number of stabilizers and commutes with the rest.
	
If $\ket{\psi_0}$ is a ground state of the Hamiltonian, then $\l^{(k)}\ket{\psi_0}$ is an eigenstate with energy $E^{(k)}$. Define the energy barrier for this particular local decomposition as
\begin{align}
\Delta_{\mathcal{D}(\bar{\l})}=\max_k(E^{(k)}-E_0),
\end{align}
where $E_0$ is the ground state energy. Then the energy barrier for the system is
\begin{align}
\Delta = \min_{\bar{\l},\mathcal{D}(\bar{\l})}\Delta_{\mathcal{D}(\bar{\l})},
\end{align}
where the minimization is taken over all local decompositions of all logical operators. Thus the system energy barrier $\Delta$ can be thought of as the minimum amount of energy that the bath must supply to perform a nontrivial logical operation.

\begin{figure}
\centering
\includegraphics{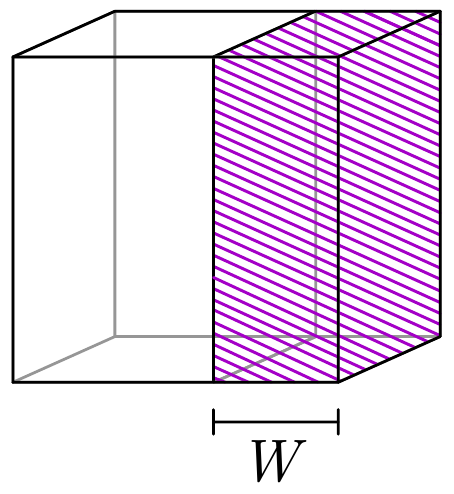}
\caption{The 1-form symmetry will be enforced within a distance $W$ from the right boundary.}
\label{fig:enforce}
\end{figure}
	
We now turn our focus to the $L\times L\times L$ 3d 3-fermion model, with the 1-form symmetry enforced within distance $W$ of the boundary as in Fig.~\ref{fig:enforce}. We want to show that the energy barrier for a boundary string operator is of order $W$.
For concreteness let the string be $S^e_\vert$, but similar constructions exist for the other strings.

\begin{figure}
\centering
\includegraphics{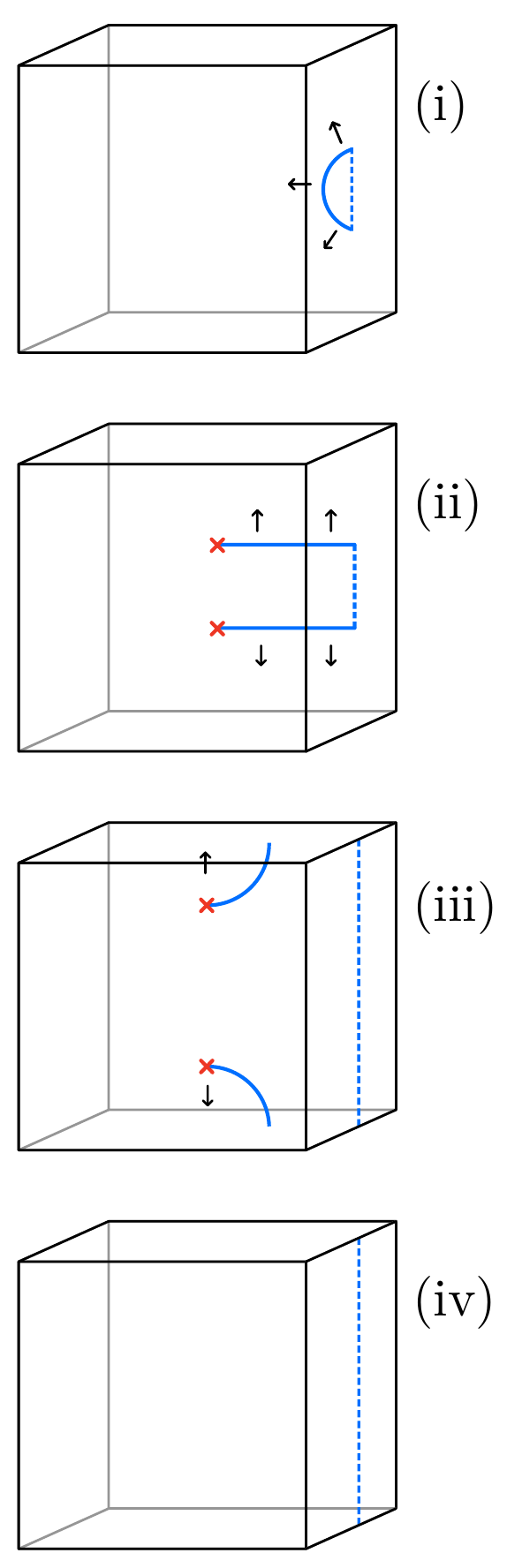}
\caption{Symmetric decomposition of a boundary string operator. Dashed blue lines represent deconfined boundary strings and solid blue lines represent bulk strings with flux. Red asterisks are point excitations. (i)~Start with a small loop near the boundary and expand it away from the boundary. (ii)~Open the loop when it is outside the symmetry-protected region. (iii)~Move the bulk anyons and flux vertically, stretching the boundary deconfined string. (iv)~Annihilate the bulk excitations, leaving behind a boundary deconfined logical string operator. Compare to Fig.~12 in Ref.~\cite{RobertsBartlett}.}
\label{fig:steps}
\end{figure}

In order to symmetrically decompose the operator, we just need to make sure the string never has an endpoint in the protected region. We start with a small loop near the boundary, as in Fig.~\ref{fig:steps}. Any part of the loop on the boundary will create no flux, while any part of the loop in the bulk will create flux.

It is possible to move the string operator using local sets of Pauli operators since if $\C$ and $\C'$ only differ in a single region, then $S^e_\C$ and $S^e_{\C'}$ only differ in the same region. We use this method to pull the edge of the loop into the unprotected region so that we are allowed to open it, breaking the 1-form symmetry. At this point (Fig.~\ref{fig:steps}~ii) we have a deconfined string operator on the boundary and two confined string operators reaching into the bulk. The excitations are two point excitations with energy cost $\sim 2$ and two flux tubes with energy cost $\sim 2W$.

We now move the confined strings in the vertical direction until they annihilate, leaving behind a deconfined logical string operator on the boundary. As $W\to \infty$ the largest energy cost comes from the flux tubes, so the symmetric energy barrier is $\Delta\sim W$.

As long as we ensure that $W$ grows without bound as we take the thermodynamic limit, this shows that 1-form symmetry protection can endow the 3d 3-fermion model with a diverging energy barrier. Furthermore, Ref.~\cite{RobertsBartlett} shows that in this type of model, a diverging energy barrier is sufficient to ensure self-correction.

Note that this means that $W$ need not scale as any particular function of $L$. We could take the thermodynamic limit in such a way that $W/L\to 0$ as long as both grow without bound, for example by taking $W\sim \log L$. Then the symmetry would be enforced in a measure zero sub-volume in the thermodynamic limit. In this sense the quantum memory only requires that the symmetry be enforced near the boundary, not in the whole bulk. This is our first signal that we are not relying on the existence of an SPT phase.

We could have let the two non-contractible directions have lengths $L_1$ and $L_2$ and not required $L_1, L_2>W$. In that case the energy barrier scales as $\Delta\sim \min\{L_1,L_2,W\}$, reproducing the above scaling when $W>L_1,L_2$. 
This scaling is reminiscent of the behavior in Ref.~\cite{RobertsBartlett}. If the vertical direction is smaller than $W$, then it is more energy efficient to first make the loop very large in the vertical direction until it split into a nontrivial boundary loop and a nontrivial bulk loop, as in Fig.~12 of~\cite{RobertsBartlett}. The bulk loop can then be moved to the unprotected region, broken, and removed.

Before moving on we will mention what happens if we enforce the symmetry everywhere in the bulk, as in the R\&B proposal \cite{RobertsBartlett}. In that case, any logical operator with a symmetric local decomposition must have a nontrivial logical action on both boundaries. Comparing to Fig.~\ref{fig:steps}, the closed string may never open, so it has to end up as a nontrivial loop on the left boundary. 

We previously said that membrane operators had to have logical actions on both boundaries. In confined Walker-Wang models, for any closed string operator there is some dual membrane operator with the same action on the ground space. This is because arbitrary products of face operators $B_f^\sigma$ and $B_f^\tau$ create open dual membranes with string operators around their perimeters.

Under a certain encoding of logical qubits 3 and 4 in the left boundary, the logical operators that can be locally decomposed are 
\begin{align}
X_1X_3 &= R^\sigma_\horiz,\qquad Z_1Z_3 = R^\tau_\vert\nn
X_2X_4 &= R^\tau_\horiz, \qquad Z_2Z_4 = R^\sigma_\vert.
\label{eqn:fullenc}
\end{align}
Note that these operators can generate any Pauli on a given qubit, but they are constrained to commute with each other. This is analogous to the 3d toric code, where 1-form symmetry-protection means that only membrane operators can be symmetrically decomposed.

The 3d 3-fermion model is a confined Walker-Wang model. All models in this family have confined anyons in the bulk and deconfined anyons on the boundary. Thus, all these models have trivial bulks with 2d topological order on the boundary. For any confined Walker-Wang model it should be possible to follow the above procedure of enforcing the 1-form symmetry within a distance $W$ of the boundary to achieve a energy barrier that scales as $\Delta\sim W$.

To close this section, we should connect to the 3D cluster state model of Raussendorf, Bravyi and Harrington, the RBH model, which was the original setting for the R\&B proposal \cite{RobertsBartlett}. Like the confined Walker-Wang models, this model is trivial in the bulk and can have boundary conditions that create topological order. When defined on the topology $T^2\times I$ with the symmetry enforced within a distance $W$ of one boundary, the RBH model protects two qubits at that boundary at nonzero temperature.

\section{Paramagnetic bulk} \label{sec:trivial}

In the previous section we saw how enforcing a 1-form symmetry on an SPT system could give rise to self correction. We also saw that the symmetry need not be enforced in the whole bulk, which leads one to wonder if the SPT nature of the bulk was really necessary. Here we present a construction inspired by Sec.~III.G of~\cite{RobertsBartlett}, in which the symmetry provides self-correction, using a non-interacting paramagnet for the bulk Hamiltonian. Since trivial paramagnets are by definition not in an SPT phase, this makes clear that the self-correction seen in these models is not an SPT effect, but rather follows purely from the 1-form symmetry.

Consider qubits placed on faces and edges of a cubic lattice. As before, let the lattice have topology $T^2\times I$. On the boundaries, only put qubits on edges. For simplicity we will refer to the sets of bulk cubes, faces, edges, and vertices as $Q, F, E$ and $V$, respectively. 
We will refer to the sets of boundary faces edges and vertices as $\partial F$, $\partial E$, and $\partial V$, respectively.

The Hamiltonian in the bulk is
\begin{align}
H_\para &= -\sum_{f\in F}X_f - \sum_{e\in E}X_e,
\end{align}
acting on all face and edge qubits. The boundary Hamiltonian is just a toric code,
\begin{align}
H_\TC &= -\sum_{v\in\partial V} A_v^\partial -\sum_{f\in\partial F}B^\partial_f,
\end{align}
where $A_f^\partial$ and $B_f^\partial$ are the normal 2d toric code terms, acting only on the boundary edge qubits. Recall there are no boundary face qubits.

\begin{figure}
\centering
\includegraphics{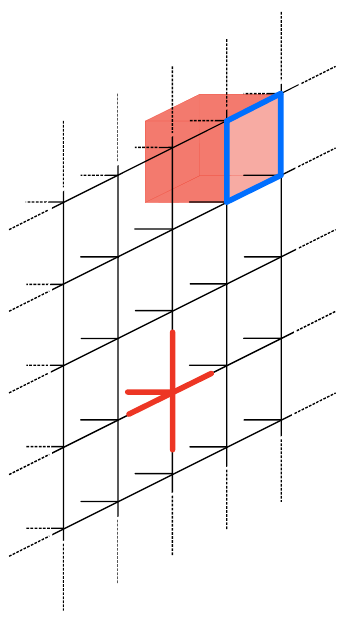}
\caption{Symmetry operators at the boundary of the lattice. The $A_v'$ operator (lower left) consists of five $X_e$ terms, while the $A_q'$ operator (upper right) has five $X_f$ terms and four $Z_e$ terms on the boundary. Restricting these terms to the boundary edges results in the terms in the boundary toric code Hamiltonian.}
\label{fig:trivbdy}
\end{figure}

The symmetry operators in the bulk are simply 
\begin{align}
A_v &= \prod_{e\in\pardag v} X_e,\qquad A_c = \prod_{f\in \partial c}X_c,
\end{align}
with one operator for each vertex and each cube. These clearly commute with the bulk Hamiltonian. We will give explicit definitions of the boundary symmetry operators but they are rather complicated so they are depicted  in Fig.~\ref{fig:trivbdy}.

On the boundary vertices, the symmetry operators are 
\begin{align}
A_v' &= \prod_{e\in\pardag v} X_e,
\end{align}
which is a five-body operator because $v$ is on the boundary.
For any cube whose boundary contains a boundary face, the symmetry operator is
\begin{align}
A_q' = \prod_{e\in \partial f^{(0)}}Z_e\prod_{f\in \partial q}X_f,\label{eqn:bdyAq}
\end{align}
where $f^{(0)}$ is the unique face in $\partial q$ on the boundary lattice. Recall the boundary faces have no qubits on them so $A'_q$ contains 5 $X$-type operators. In addition, it is dressed by a 4-body $Z$-type term on boundary qubits. See Fig.~\ref{fig:trivbdy} for illustrations.

The group generated by $A_q$, $A_v$, $A_q'$, and $A_v'$ is a 1-form symmetry because elements of the group act on codimension-1 objects. Elements generated by $A_v$ and $A_v'$ consist of $X$ operators on sets of edges forming dual membranes. These dual membranes may terminate at the lattice boundary.
Elements generated by $A_q$ and $A_q'$ consist of $X$ operators acting on sets of faces forming direct membranes. A membrane $\cal{M}$ may terminate at the lattice boundary if it is decorated by $Z$ operators on the edges that make up $\partial\cal{M}$. This decoration comes from the decoration in Eqn.~\ref{eqn:bdyAq}.

Unsurprisingly, the topological order lives in the 2-dimensional toric code at the lattice boundary. The logical operators are, as always, either direct strings of $Z$ operators or dual strings of $X$ operators.

\begin{figure}
\centering
\includegraphics{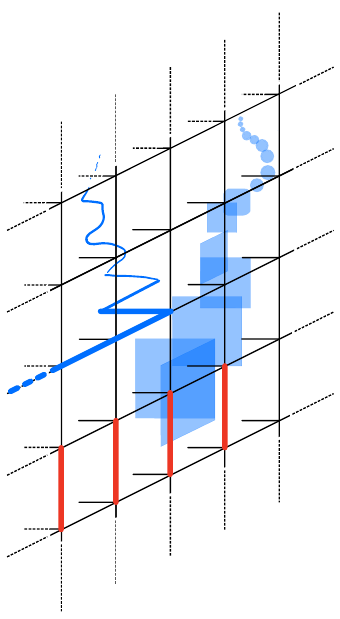}
\caption{In order to decompose the boundary logical operators in the presence of the 1-form symmetry, we need to connect boundary strings to bulk strings. The $Z$-type boundary string can simply be connected to a $Z$-type bulk string defined on edges, so that the entire string has no endpoints. The $X$-type boundary string anti-commutes with two $A_q'$ operators at its endpoints. These can also be seen as the endpoints of a bulk dual $Z$-string, so a combination of a boundary dual $X$-string on edges and a bulk $Z$-string on faces commutes with the symmetry. In both cases the bulk string creates excitations, and is linearly confined.}
\label{fig:symdecomp}
\end{figure}

Neither of these strings can be symmetrically decomposed using open boundary strings, the way they would be decomposed in a 2d toric code. Open $Z$-strings anticommute with $A_v$ operators at their endpoints. This can be fixed by pairing with a string of $Z$ operators through the bulk. Similarly, open dual $X$-strings anticommute with $A_q$ operators at their endpoints and must be paired with dual $Z$-strings through the bulk. See Fig.~\ref{fig:symdecomp} for these local symmetric decompositions.

Direct and dual $Z$-strings in the bulk commute with the 1-form symmetry because they intersect every cube or vertex term twice. However, they have linear energy cost because they anticommute with the paramagnet Hamiltonian. We can once again call them fluxes. The 1-form symmetries enforce that the fluxes can only end on the endpoints of open line operators on the toric code boundary or in regions where the symmetry is not enforced.

We find ourselves in a position similar to the 3d 3-fermion model, where boundary anyons are confined by flux strings in the bulk. Thus, we can decompose logical operators using the steps in Fig.~\ref{fig:steps}. Once again we find that the topologically nontrivial operators that can be symmetrically decomposed into strings that intersect the boundary but end deep in the bulk. Bulk strings are linearly confined, so the symmetric energy barrier for this system diverges.

\section{Discussion}

The  purpose of this paper have been to relate the R\&B construction to other existing models and to determine what aspects of the construction are most important for achieving self-correction. We showed that enforcing the 1-form symmetry in any confined abelian Walker-Wang model results in self-correction. Ref.~\cite{RobertsBartlett} conjectured that this might be possible, and indeed it is. In the process we discovered that it suffices to enforce the symmetry in a measure zero sub-volume of the system. This then led us to conjecture that it might be possible to achieve self correction with a paramagnetic bulk, and indeed, we were able to demonstrate this by explicit construction. This strongly suggests that it is the 1-form symmetry that does the heavy lifting, and any `exotic' nature of the bulk is optional. 

The approaches we have discussed achieve self-correction by giving anyons effective long-range interactions, by tying them to confined bulk flux strings. Thus they could be compared to earlier literature that also tried to utilize long-range interactions to achieve self correction ~\cite{Hamma2009, Chesi2010, Pedrocchi2011, Hutter2012, Wootton2013, Becker2013, Pedrocchi2013, Hutter2014}.
Those attempts were limited by requirements for unbounded operator strength and/or instability to perturbation~\cite{Brown2016, LandonCardinal2015}. If we enforce the 1-form symmetry by endowing certain terms in the Hamiltonian with infinite coupling constants, then the R\&B proposal (and our extensions thereof) suffer from the same limitations. 

Since we do not use any exotic bulk properties, we should ask if we can improve the construction by using a more interesting bulk.
A key direction for future work is  whether 1-form symmetry can naturally emerge in the dynamics of some quantum system. In this context, R\&B conjectured that the 3d gauge color code~\cite{BombinGauge} (gcc) might realize an emergent 1-form symmetry (including at non-zero temperature). However, the proof or disproof of this conjecture remains an open problem \cite{Kubica2018}, as does identification of other potential platforms for emergent 1-form symmetry at non-zero temperature. An alternative direction to pursue might be to seek quantum computational architectures where 1-form symmetry may be natively enforced, for instance through single shot error correction \cite{Roberts2017, BombinSingleShot}. 

The appeal of the 3d gauge color code is that the flux tubes do not end in the bulk.
The reason for the 1-form symmetry in the bulk in the R\&B proposal and in this paper was to prevent the flux tubes from terminating, so the 3d gcc would not need this constraint.
The 3d toric code is a useful point of comparison for the 3d gcc.
Like the 3d gcc, the 3d toric code has flux tubes that do not terminate in the bulk.
If it were possible to couple the endpoints of these flux tubes to boundary anyons this may result in some nonzero-temperature stability, even without higher-form symmetry enforcement. The difficulty of understanding emergent higher-form symmetry can also be seen in the 3d toric code, which has an emergent 1-form symmetry at $T=0$ but not at nonzero temperatures. 

We could consider enforcing a 1-form symmetry in the bulk of a pure 3d toric code, with no boundary anyons.
This prevents the creation of point excitations, so the stringlike operators cannot be locally decomposed. 
In the case of the 3d toric code this does promote the code to be self-correcting. The cost is that some logical operators now have no symmetric local decomposition. 

Lastly, we wonder what ingredients can be added to these models to improve the finite temperature behavior. Possibilities could include a mix of 3-dimensional and 2-dimensional topological order or boundaries between different phases instead of boundaries with the vacuum. It might be useful to use the process of welding~\cite{MichnickiPowerLaw}, which is known to create a code with a power-law energy barrier at $T=0$. Furthermore, fracton phases (see \cite{fractonarcmp} for a review) give access to new kinds of bulk order that could be also useful in this quest. We leave these explorations to future work. 

{\bf Acknowledgements}: This work was supported by the U.S. Department of Energy, Office of Science, Basic Energy Sciences, under Award \# DE-SC0021346
\bibliography{bob}

\end{document}